# Realization of polytype heterostructures via delicate structural transitions from a doped-Mott insulator

Yanyan Geng[1,2,+], Manyu Wang[1,2,+], Shumin Meng[1,2], Shuo Mi[1,2], Chang Li[1,2], Huiji Hu[1,2], Jianfeng Guo[3], Rui Xu[1,2], Fei Pang[1,2], Wei Ji[1,2], Weichang Zhou[4*], and Zhihai Cheng[1,2,*]

[1]Key Laboratory of Quantum State Construction and Manipulation (Ministry of Education), School of Physics, Renmin University of China, Beijing 100872, China

[2]Beijing Key Laboratory of Optoelectronic Functional Materials & Micro-nano Devices, School of Physics, Renmin University of China, Beijing 100872, China

[3]Beijing National Laboratory for Condensed Matter Physics, Institute of Physics, Chinese Academy of Sciences, Beijing 100190, China

[4]Key Laboratory of Low-dimensional Quantum Structures and Quantum Control of Ministry of Education, School of Physics and Electronics, Hunan Research Center of the Basic Discipline for Quantum Effects and Quantum Technologies, Hunan Normal University, Changsha 410081, China

**Abstract:** Transition metal dichalcogenides (TMDs) host multiple competing structural and electronic phases, making them an ideal platform for constructing polytype heterostructures with emergent quantum properties. However, controlling phase transitions to form diverse heterostructures inside a single crystal remains challenging. Here, we realize vertical/lateral polytype heterostructures in a hole-doped Mott insulator via thermal-annealing-induced structural transitions. Raman spectroscopy, atomic force microscopy (AFM) and scanning Kelvin probe force microscopy (SKPM) confirm the coexistence of T-H polytype heterostructures. Atomic-scale scanning tunneling microscopy/spectroscopy (STM/STS) measurements reveal the transparent effect in 1H/1T vertical heterostructures, where the charge density wave (CDW) of the underlying 1T-layer superposes on the top 1H-layer under positive bias. By systematically comparing 1T/1H and 1T/1T interfaces, we demonstrate that the metallic 1H-layer imposes a Coulomb screening effect on the 1T-layer, suppressing the formation of CDW domain walls and forming more ordered electronic states. These results clarify the interfacial coupling between distinct quantum many-body phases and establish a controllable pathway for constructing two-dimensional polytype heterostructures with tunable electronic properties.

[+]These authors contributed equally: Yanyan Geng, Manyu Wang

[*]Email: wchangzhou@hunnu.edu.cn, zhihaicheng@ruc.edu.cn

**Introduction**

Layered transition metal dichalcogenides (TMDs) have emerged as an ideal material platform for exploring correlated electronic behaviors due to their rich and tunable electronic states, including charge density waves (CDW) [1-3], quantum spin liquids [4], superconductivity [5-7], and Mott insulators [8-10]. Their intrinsic two-dimensional nature allows van der Waals polytype heterostructures to be constructed by combining monolayers with distinct structural phases, while largely preserving the individual properties of each layer [11-14]. More importantly, interlayer interactions can strongly intertwine the characteristics of different components, giving rise to emergent electronic states that surpass the intrinsic behavior of the individual materials, such as topological nodal-point superconductivity [15], interfacial Kondo physics [16], charged interlayer excitons [17], and interlayer charge transfer [18,19]. These phenomena highlight the substantial potential of polytype heterostructures for engineering correlated quantum states. However, conventional methods for fabricating such heterostructures typically rely on artificial stacking or epitaxial growth [20-23], suffer from inefficiency, susceptibility to impurities, and defects. Achieving controllable, high-quality, and phase-pure polytype heterostructures within a single crystal remains challenging.

$TaS_2$, as a prototypical TMDs material, serves as an exemplary material for exploring polytype heterostructures owing to its rich crystal structures and phase-dependent electronic correlation properties [24-27]. Specifically, $1T$-$TaS_2$ exhibits the Mott insulating ground state [28,29], whereas $1H$-$TaS_2$ demonstrates metallic and superconducting behavior at low temperatures [30,31]. The atomic-scale integration of these two electronically distinct phases offers an ideal platform for manipulating electron correlation and interlayer coupling. A series of emergent phenomena have been observed in such systems, including artificial heavy-fermion behavior [32], chiral superconductivity [33], two-component nematic superconductivity [34], and Kondo-like responses [35,36]. Notably, external stimuli such as laser, temperature, or electric fields can trigger a structural phase transition from $1T$-$TaS_2$ to $1H$-$TaS_2$ [37-39], enabling the in situ construction of T-H polytype heterostructures within a single crystal. Furthermore, hole doping can drive the system into a critical regime of competing phases, opening new avenues for heterostructure engineering and manipulation of exotic electronic states. However, the direct realization of stable T-H polytype heterostructures starting from a hole-doped Mott insulator has been rarely reported.

In this work, we successfully realize diverse polytype heterostructures in hole-doped $1T$-TaS$_2$ crystals through thermally induced structural phase transitions. The formation of both vertical and lateral T-H heterostructures is confirmed by Raman spectroscopy, Atomic Force Microscope (AFM) and scanning Kelvin probe force microscopy (SKPM). High-resolution STM imaging further uncovers a distinct electronic transparency effect at 1H/1T interfaces under positive bias voltage, where the $\sqrt{13}\times\sqrt{13}$ CDW modulation of the underlying 1T-layer can be detected through the top metallic 1H-layer. Comparative investigations of 1T/1H and 1T/1T heterostructures reveal that the 1T-layer is subjected to Coulomb screening by the metallic 1H-layer, which reduces the intralayer electron correlations and suppress the formation of phase domain walls. This work not only achieves multi-dimensional polytype heterostructures in a single crystal, but also provides a new platform for investigating proximity effects between Mott insulating, metallic, and superconducting phases in two-dimensional materials.

**Results and discussion**

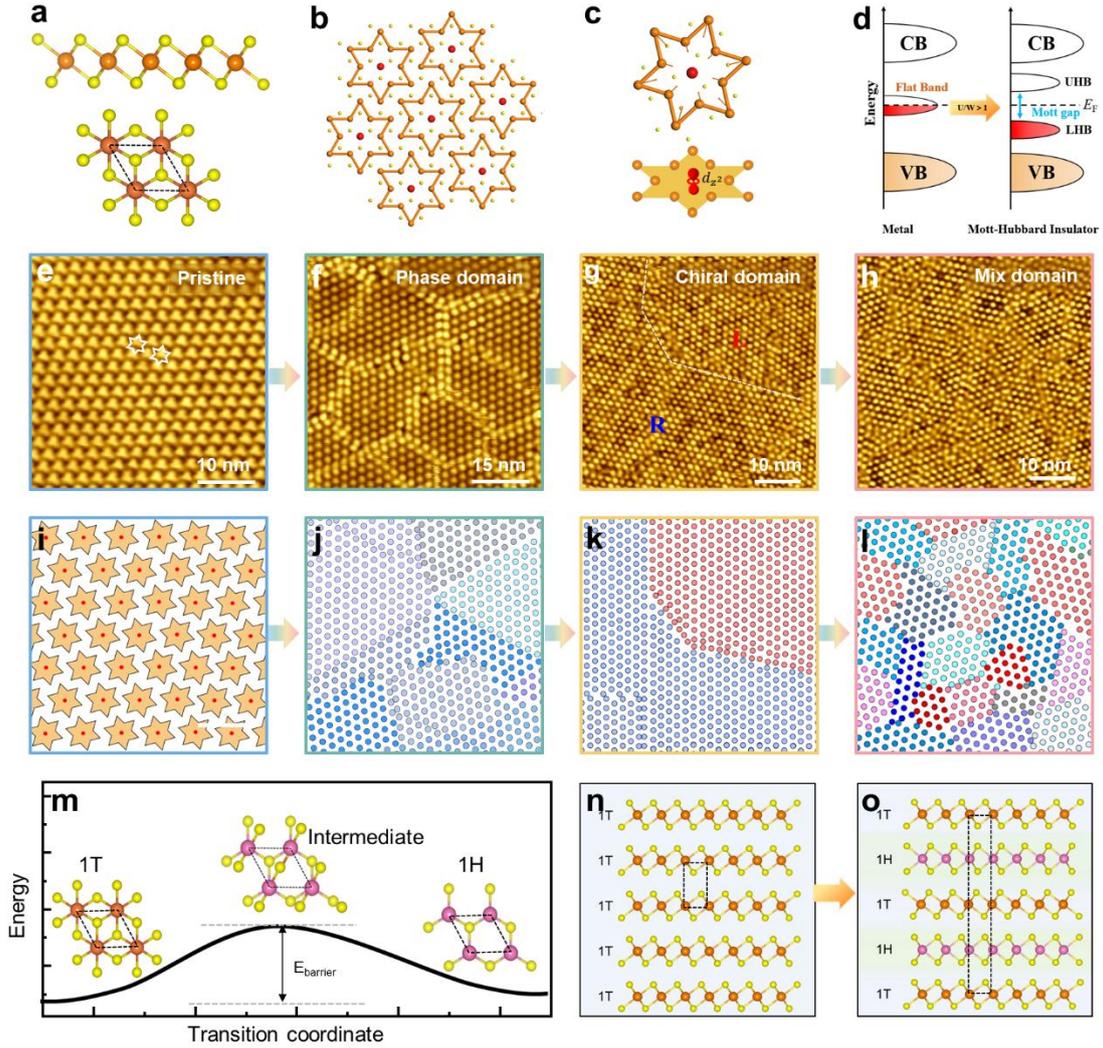

**Figure 1. Realization of polytype heterostructures via the doped-Mott insulator.** (a) Atomic structure model of monolayer 1$T$-TaS$_2$. (b) Schematic of the √13×√13 Star-of-David (SoD) cluster superstructure. (c) Atomic models of the SoD cluster with the localized $d_{z^2}$ orbital (unpaired electron) of central Ta atom. (d) Schematic band structures of correlated CDW state, where the half-filled flat band splits into UHB and LHB bands via electron correlations. (e-l) STM images and corresponding structure models of the pristine, low-hole-doped (x=~0.77%), medium-hole-doped (x=~2.5%) and high-hole-doped (x=~5%) TaS$_2$. As the hole doping concentration increases, phase domains (f,g), chiral domains (g,k), and mixed domains (h,l) progressively appear. (m) Schematic energy landscape and top-view lattice structures of the 1T-phase (left), intermediate state (middle), and 1H-phase (right). $E_{barrier}$ denotes the energy barrier between the 1T- and 1H- phases. (n,o) Schematic illustration of the transition from the 1T phase (n) to the 4Hb-phase (o) in the doped-Mott insulator. Scanning parameter: (e) $V$=0.5 V, $I$=100 pA; (f-h) $V$=1.0 V, $I$=120 pA.

The 1$T$-TaS$_2$ is a prototypical platform for studying Mott physics due to its unique correlated electron ground state. Its layered structure consists of Ta atoms sandwiched between two S layers, forming an octahedral 1T coordination, as shown in Fig. 1(a). At low temperature, the ground CDW state of 1$T$-TaS$_2$ is characterized by the commensurate √13×√13 superlattice of Star of David (SoD) clusters, as depicted in Figs. 1(b) and 1(c). Each SoD contains 13 Ta 5d electrons, the 12 electrons of the outer Ta atoms pair and form six occupied CDW bands, and leave one unpaired electron of the central Ta atom in a half-filled flat band. This half-filled band further splits into upper and lower Hubbard bands (UHB and LHB) due to the strong electron-electron correlations, resulting in a Mott insulator [Fig. 1(d)]. This Mott insulating state is highly sensitive to carrier perturbations, making 1$T$-TaS$_2$ an ideal material for manipulating the exotic quantum states.

Hole doping can effectively modify the flat-band filling factor and disturbs the CCDW order of 1$T$-TaS$_2$, triggering a sequence of novel emergent quantum states [27]. Real-space STM images and structural models reveal that hole doping disrupts long-range CCDW order, gradually inducing the formation of phase domains, chiral domains, and mixed phase/chiral domains, as shown in Figs. 1(e)-1(l). Moreover, the introduction of holes weakens the intralayer and interlayer interactions, pushing the system into a critical regime where multiple electronic phases compete with similar energies. Consequently, hole-doped 1$T$-TaS$_2$ not only disrupts the original Mott insulating state but also places the system in a tunable correlated phase space, laying the foundation for realizing the polytype heterostructures.

Previous studies have shown that the energy difference between the 1T- and 1H-phases is very small, as shown in Fig. 1(m). The phase transition between 1T-phase and 1H-phase can be achieved by various external means such as laser excitation, STM-tip pulse, and thermal annealing [37-39]. The weakening of electron correlations induced by hole doping renders the system more easily to overcome the energy barrier between the 1T-phase and the 1H-phase. Thermal annealing can further promote delicate structural transitions, enabling the transition from a doped Mott insulator to polytype heterostructures, as illustrated in Figs. 1(n) and 1(o). These polytype heterostructures not only retain the electron correlations of the 1T-phase but also introduce the metallic/superconducting state of the 1H-phase. It provides a unique platform for studying the competition, coupling, and proximity effects between electron correlations, CDW, and superconductivity. Such a strategy of synergistic control through "doping and interface" to achieve structural transition significantly enriches the pathways for electronic state manipulation.

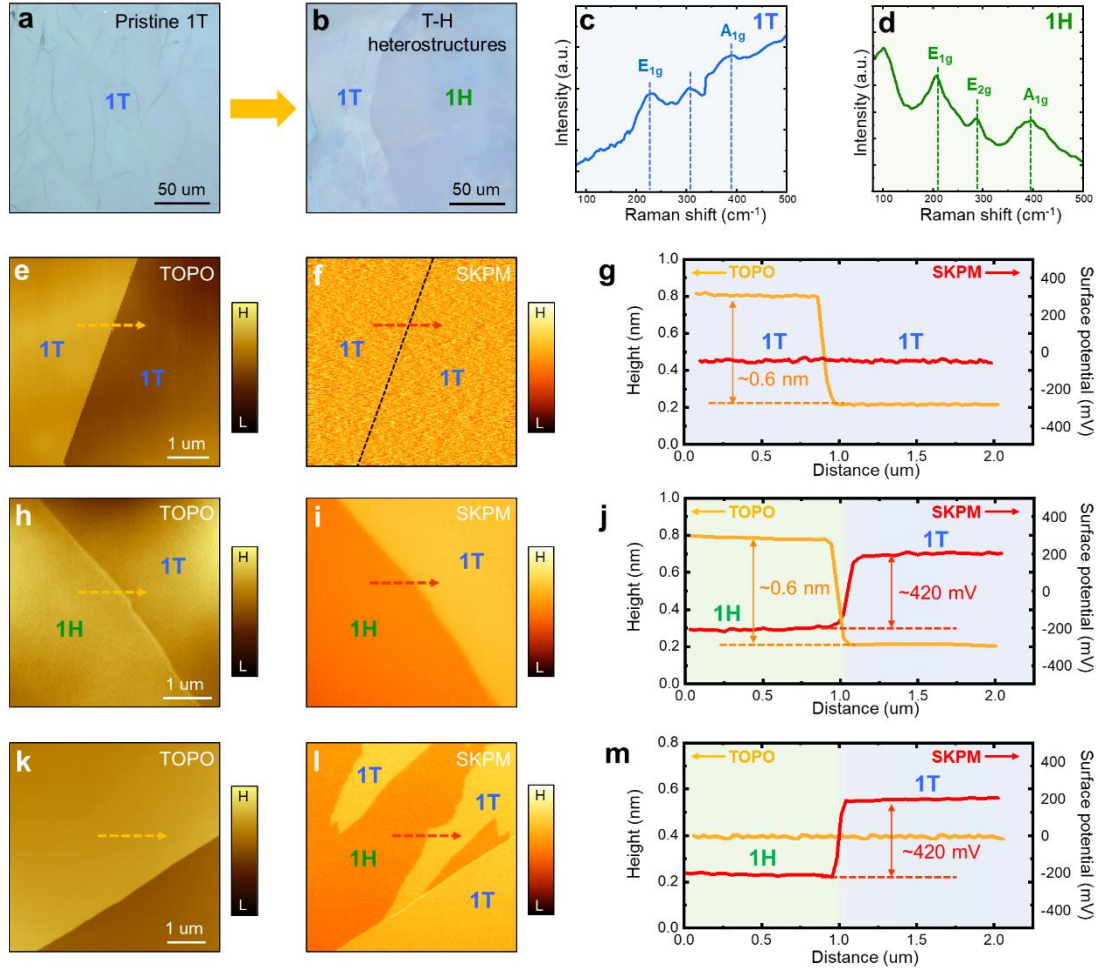

**Figure 2. Raman and SKPM characterization of the polytype heterostructures in low-hole-doped TaS$_2$.** (a,b) Typical optical images of the TaS$_2$ before (a) and after (b) the formation of polytype heterostructures. (c,d) Raman spectra acquired from the 1T-phase (c) and 1H-phase (d) regions indicated in (a). Characteristic Raman modes of the 1T- and 1H-phases are marked with green and blue dashed lines, respectively. (e,f) AFM topography (e) and corresponding SKPM surface potential (f) images of a pristine 1T structure. (g) Line profiles extracted along the dashed lines in (e) and (f). The step height between adjacent 1T-layers is ~0.6 nm, with no significant surface potential difference observed. (h,i) AFM topography (h) and SKPM surface potential (i) images of vertical 1T-1H heterostructures. (j) Line profiles along the dashed lines in (h) and (i). The step height and surface potential difference between the 1T- and 1H-layers are ~0.6 nm and ~420 mV, respectively. (k,l) AFM topography (k) and SKPM surface potential (l) images of lateral 1T-1H heterostructures. (m) Line profiles along the dashed lines in (k) and (l). The surface potential difference between the 1T-domain and 1H-domain is ~420 mV.

For low-hole-doped 1$T$-TaS$_2$, local lattice reconstruction and phase transitions can be induced when the annealing temperature rises above ~450 °C. Optical microscope images before and after annealing show significant color differences, which indicate the transition from 1T to T-H polytype heterostructures, as shown in Figs. 2(a) and 2(b). To further confirm this structural transition, we perform Raman spectroscopy on different regions of Fig. 2(b). In the 1T-phase region [Fig. 2(c)], the Raman spectrum shows three characteristic peaks at approximately 240, 310, and 380 cm$^{-1}$, corresponding to the $E_{1g}$ mode, the folded phonon mode related to the CDW, and the $A_{1g}$ mode, respectively. In contrast, the Raman spectrum of 1H-phase exhibits the characteristic peaks at approximately 210, 285, and 400 cm$^{-1}$ [Fig. 2(d)], corresponding to the $E_{1g}$, $E_{2g}$, and $A_{1g}$ modes. The Raman spectra confirms that thermal annealing drives the structural reconstruction from 1T-phase to 1H-phase, achieving the T-H polytype heterostructures.

The AFM topography and SKPM surface potential images of 1T-phase and T-H heterostructures are further given in Figs. 2(e)-2(m) and Fig. S1. For the pristine 1T-phase region [Figs. 2(e)-2(g)], the AFM topography shows a uniform layered structure, with a step height of ~0.6 nm. The corresponding SKPM image and line profile show no observable surface potential difference between different 1T-layers, reflecting its uniform Mott insulating properties. In the T-H heterostructure, the 1T- and 1H-layers can be easily distinguished in the SKPM images by their relatively higher and lower surface potentials, respectively.

In the vertical 1T/H heterostructure region [Figs. 2(h)-2(j)], the surface potential difference between 1T-layer and 1H-layer are ~420 mV. Notably, this potential difference is higher than the values measured in intrinsic 4Hb-TaS$_2$ and Se-doped 4Hb-TaS$_2$ [40]. We suggest that hole doping decrease the work function of the 1T-phase, thereby increasing the potential difference relative to the 1H-phase. In contrast, isoelectronic Se doping primarily increases the interlayer spacing and weakens interlayer coupling. This reduced coupling suppresses charge transfer, which relatively increase the work function of the 1T-phase. Moreover, the SKPM results further demonstrate that a potential difference of ~420 mV is also present at the in-plane domain boundaries, confirming the lateral T-H heterojunctions. The realization of such multi-dimensional polytype heterostructures provides a new pathway for directly studying both vertical and lateral proximity effects between Mott insulators and metals/superconductors.

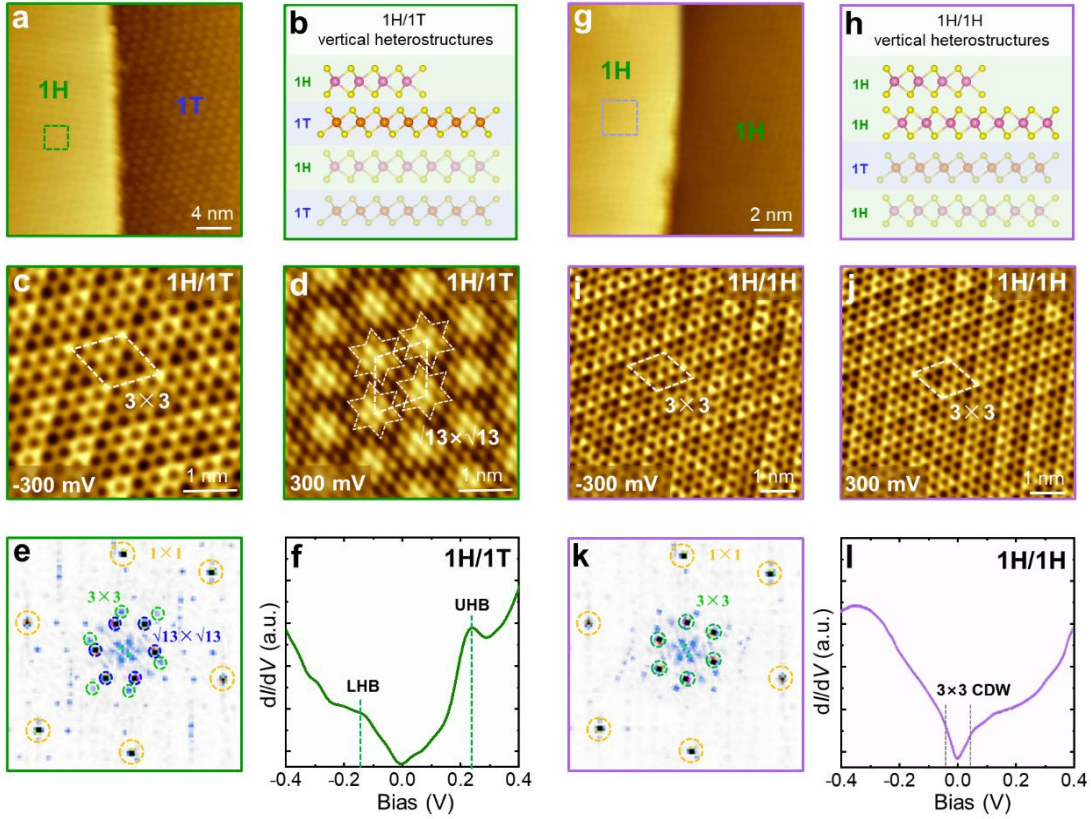

**Figure 3. Structure and electronic properties of vertical 1H/1T and 1T/1T heterostructures.** (a,b) STM image and structure model of the 1H/1T heterostructure. (c,d) STM images acquired from the same region of the 1H/1T surface under negative (c) and positive (d) bias. At negative bias, the 3×3 CDW of the top 1H-layer is visible, whereas positive bias reveals the √13×√13 CDW of the underlying 1T-layer. (e) The Fast Fourier transform (FFT) image of (d). The green and blue circles correspond to the 3×3 CDW and √13 ×√13 CDW wavevectors, respectively. (f) The d$I$/d$V$ spectra taken on the 1H/1T surface. In addition to the quasi-3×3 CDW pseudogap of 1H-layer at the Fermi level, the UHB and LHB features of the 1T-layer are also observed. (g,h) STM image and structure model of the vertical 1H/1H heterostructure. (i,j) STM images taken on the same region of 1H/1H surface under negative (i) and positive (j) bias. Only the 3×3 CDW pattern of the 1H/1H surface is visible. (k) The FFT image of (j). (l) The d$I$/d$V$ spectra of the 1H/1H surface, showing only the quasi-3×3 CDW pseudogap. Scanning parameter: (a,c,g,i) $V$=-0.3 V, $I$=-200 pA; (d,j) $V$=0.3 V, $I$=200 pA.

Figure 3 presents the atomic-scale STM characterization of the vertical 1H/1T and 1H/1H polytype heterostructures, revealing the distinct interfacial electronic modulations arising from different stacking configurations. Figures 3(a)-3(f) display the structure and electronic properties of vertical 1H/1T heterostructure. Under negative bias (occupied state), the 1H/1T surface exhibits the intrinsic 3×3 CDW superlattice of the top 1H-layer, as marked by the green rhombus in Fig. 3(c). In contrast, under positive bias (unoccupied state), a pronounced electron transparency effect is observed, wherein the √13×√13 periodic modulation of the underlying 1T-layer is superimposed upon the 3×3 pattern of 1H-layer [Fig. 3(d)]. The Fast Fourier transform (FFT) analysis further confirms the coexistence of these two periodicities, where the green and blue dashed circles correspond to the 3×3 and √13×√13 wavevectors, respectively [Fig. 3(e) and Fig. S2].

This bias-dependent behavior indicates that, under positive bias, electrons from the STM tip can tunnel through the top 1H-layer into the unoccupied states of the underlying 1T-layer. In contrast, under negative bias, the occupied states of the 1H-layer form an energy barrier, causing tunneling current to mainly reflect the electronic states of the 1H-layer [Fig. S3]. However, this simple tunneling-decay model cannot fully account for why the CDW modulation of the 1T-layer remains clearly visible despite the finite thickness of the 1H-layer and the presence of the van der Waals gap. To further deeply understand the underlying mechanism, the d$I$/d$V$ spectroscopy on the 1H/1T region is performed [Fig. 3(f)]. In addition to the quasi-3×3 CDW pseudogap near the Fermi level from the 1H-layer, the spectra also clearly reveal the characteristic LHB and UHB features of the 1T-phase. These results demonstrate that interlayer electronic coupling enables the correlated electronic states of the 1T-layer to effectively influence the local density of states in the 1H-layer, thereby further enhancing the visibility of the CDW modulation of the 1T-layer in STM images.

We further conducted comparative studies on vertical 1H/1H heterostructures, as depicted in Figs. 3(g)-3(l). STM images acquired under both negative and positive biases [Figs. 3(i) and 3(j)] display only the intrinsic 3×3 CDW pattern of the 1H-phase. The d$I$/d$V$ spectrum of the 1H/1H region similarly shows only the quasi-3×3 CDW pseudogap without any other electronic features, as shown in Fig. 3(l). This comparison confirms that the additional √13×√13 modulation arises from interfacial electronic coupling unique to the polytype 1H/1T interface, providing evidence for interface-driven electronic modulation in correlated layered systems.

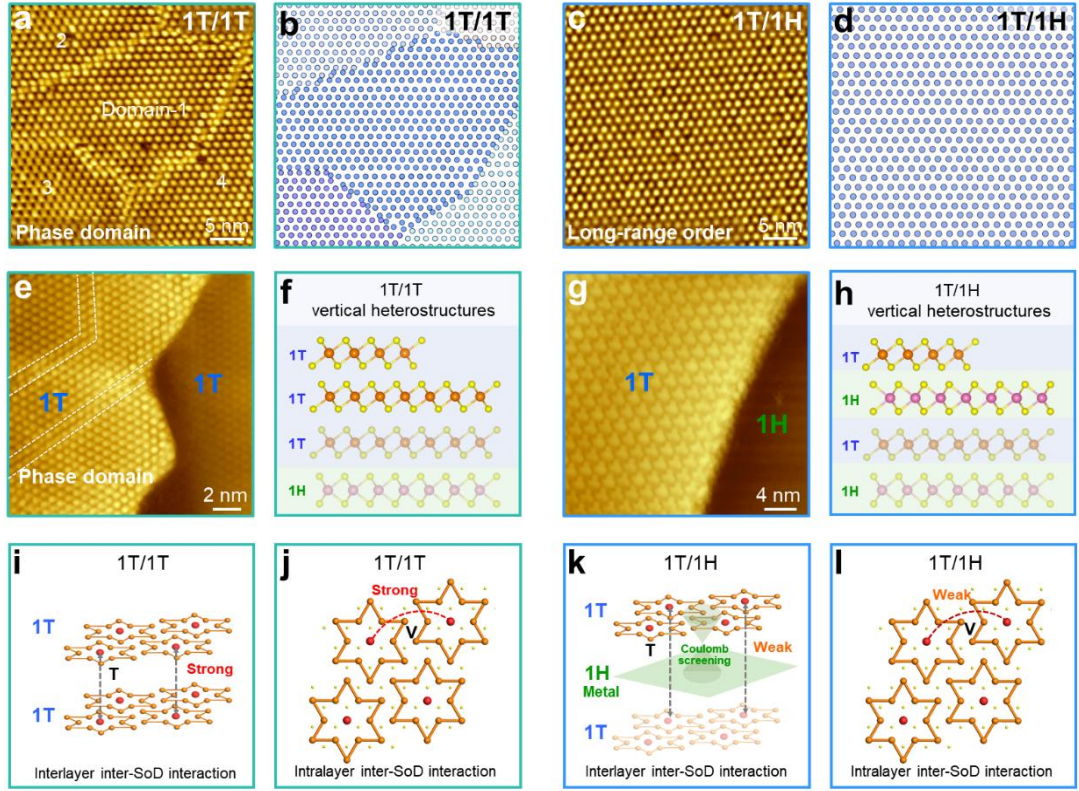

**Figure 4. Structure properties of vertical 1T/1H heterostructures.** (a,b) STM image and schematic diagram of the 1T/1T surface, with the phase domain walls clearly visible. (c,d) STM image and schematic of the 1T/1H surface, where the phase domain walls are suppressed. (e,f) STM image and schematic diagram of the 1T/1T step edge. Phase domain walls in the top 1T-layer are indicated by white dashed lines. (g,h) STM image and schematic diagram of the 1T/1H step edge. (i,j) Schematics illustrating interlayer (T) and intralayer (V) inter-SoD interactions in the 1T/1T configuration. The 1T/1T configuration is under strong interlayer and intralayer inter-SoD interactions. (k,l) Schematic of the interlayer (T) and intralayer (V) inter-SoD interactions of the 1T/1H configuration. The Coulomb screening effect from the metallic 1H-layer weakens the intralayer inter-SoD interactions, resulting in weak interlayer and intralayer inter-SoD interactions in 1T/1H configuration. Scanning parameter: (a,c) $V$=1.0 V, $I$=100 pA; (e,g) $V$=0.5 V, $I$=100 pA.

Figure 4 further presents the structure properties of the vertical 1T/1T and 1T/1H polytype heterostructures, revealing the regulation mechanisms of different stacking configurations on the CDW domain structure and internal interactions of 1T-phase. As illustrated in Figs. 4(a) and 4(b), the 1T/1T surface exhibits the typical phase domain network, consistent with the bulk low-hole-doped 1$T$-TaS$_2$ sample. In contrast, the STM image of the 1T/1H surface displays a uniform and long-range ordered structure with strongly suppressed domain walls [Figs. 4(c), 4(d) and Fig. S4]. This indicates that the underlying metal 1H-layer exerts a significant modulating effect on the CDW domains of the 1T-layer. To further validate the influence of different stacking configurations on the electronic order of the 1T-layer, the step edges of 1T/1T and 1T/1H configurations are displayed in Figs. 4(e)-4(h) and Fig. S5. In the 1T/1T step region [Figs. 4(e) and 4(f)], the upper 1T-layer still exhibits a dense network of phase domain walls, marked by white dashed lines. In contrast, the domain structure is similarly weakened at the 1T/1H step region [Figs. 4(g) and 4(h)], consistent with the observations in Fig. 4(c). This agreement further confirms that the metallic 1H-layer effectively modulates the local electronic structure of the adjacent 1T-layer.

Figures 4(i)-4(l) illustrate the microscopic mechanism underlying the distinct behaviors of these heterostructures. In the 1T/1T configuration [Figs. 4(i) and 4(j)], both the interlayer inter-SoD interaction (T) and intralayer inter-SoD interaction (V) remain relatively strong [27]. As a result, the system stabilizes a strong correlated Mott-CDW state, thereby giving rise to abundant phase domains. Conversely, in the 1T/1H heterostructure [Figs. 4(k) and 4(l)], the 1H-layer weakens the interlayer coupling and reduces the interlayer inter-SoD interaction. Moreover, the free charge carriers within the metallic 1H-layer exert the coulomb screening effect on 1T-layer. This Coulomb screening effect diminishes the intralayer inter-SoD interaction within the 1T-layer, ultimately manifesting as the disappearance of CDW domain structures and the emergence of a long-range ordered electron distribution. These findings demonstrate that polytype stacking configurations offer an effective degree of freedom for tuning the correlation strength and CDW domain configuration in $1T$-TaS$_2$. By designing 1H/1T and 1T/1T stacking sequences within a single crystal, synergistic control over electronic correlation, interlayer coupling, and phase domain structures can be achieved in single-crystal systems.

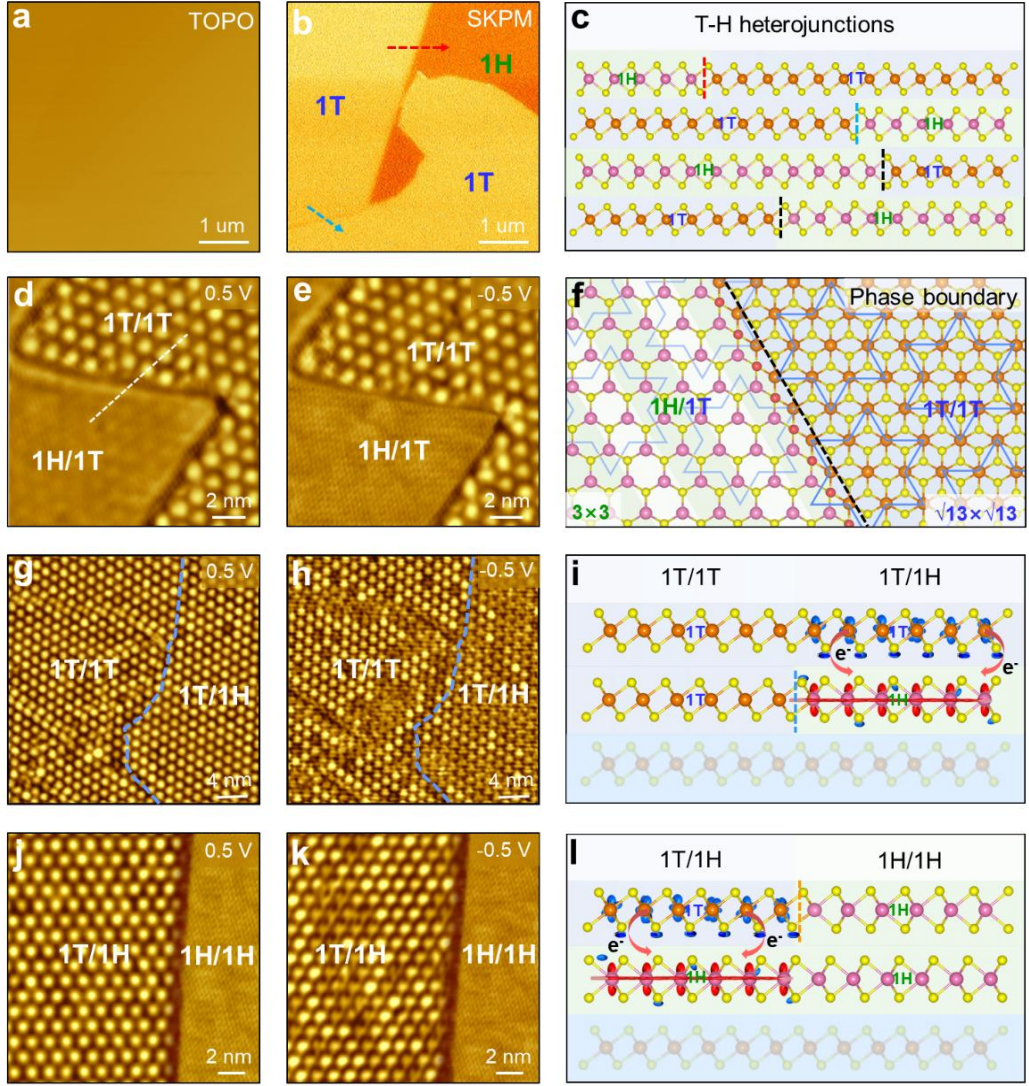

**Figure 5. Electronic properties of polytype lateral heterojunctions.** (a,b) AFM topography (a) and SKPM surface potential (b) images of lateral 1T-1T and 1T-1H heterojunctions. (c) Lateral view of the polytype vertical/lateral T-H heterostructures. (d,e) STM images of lateral 1H/1T and 1T/1T heterojunctions taken with positive (c) and negative (d) bias, with the SoD is distinctly observable under positive bias. The white dotted line indicates the directions of the SoD in the 1H/1T and 1T/1T. (f) Schematic of the lateral 1H/1T and 1T/1T heterojunctions. The sharing S atoms at the phase boundary are highlighted by red and the black dotted line indicates the phase boundary in the 1H/1T and 1T/1T. (g,h) STM images of lateral 1T/1T and 1T/1H heterojunctions taken with positive (g) and negative (h) bias. (i) Schematic of the lateral 1T/1T and 1T/1H heterojunctions. The charge transfers at the 1T/1H heterointerface contributes the apparent difference of 1T-layer at occupied and unoccupied states. Blue and red represent the electron depletion and accumulation regions. (j,k) STM images of lateral 1T/1H and 1H/1H heterojunctions taken with positive (j) and negative (k) bias. (l) Schematic of the lateral 1T/1H and 1H/1H heterojunctions.

The interface electronic properties in lateral heterojunctions are further systematically investigated, as shown in Fig. 5. Large-scale AFM topography and SKPM surface potential images reveal the micrometer-scale lateral 1T-1T and 1T-1H heterojunctions in TaS$_2$, as shown in Figs. 5(a)-5(c). Atomic-scale STM characterization further reveals rich electronic behaviors at these lateral heterojunctions [Figs. 5(d)-5(l)]. At the 1H/1T and 1T/1T lateral heterojunctions, the SoD superlattice can be resolved in 1H/1T region under positive bias, supporting the universality of the electronic transparency effect identified in the vertical 1H/1T heterostructures. Notably, the orientations of the SoDs in the 1H/1T and 1T/1T regions exhibit consistent orientation, as marked by the white dashed lines in Fig. 5(d). The structural model further shows that the top S-atom arrangement remains nearly identical across the lateral phase boundary between the 1H- and 1T-phases, as illustrated in Fig. 5(f).

In contrast, the 1T/1T and 1T/1H lateral heterojunctions display markedly different electronic properties. The CDW domain walls are observed in the 1T/1T region, whereas the domain structure is strongly suppressed in the 1T/1H region [Fig. 5(g)], confirming the pronounced influence of the metallic 1H-layer on the CDW order of the 1T-layer. More intriguingly, under positive bias (unoccupied states), all SoD units in both 1T/1T and 1T/1H regions exhibit similar shapes and intensities. However, under negative bias (occupied states), the number of bright SoDs in the 1T/1H region is significantly reduced compared to the uniformly distributed SoDs in the 1T/1T region, displaying alternating bright and dark SoD patterns [Fig. 5(h), Figs. S6 and S7]. This difference can be attributed to the interlayer electron transfer at the 1T/1H heterointerface, as presented in Fig. 5(i). The STM images of the 1T/1H and 1H/1H lateral heterojunctions further confirm that such charge transfer leads to distinct electronic structures in the occupied and unoccupied states of the 1T-layer [Figs. 5(j)-5(l)]. This behavior is consistent with the bright-dark SoD patterns induced by partial charge transfer previously observed in Se-doped 4Hb-TaS$_2$ [40].

Our findings reveal that doping-driven structural transitions can simultaneously construct both vertical and lateral polytype heterostructures within a single-crystal system. We have identified universal electronic transparency effects and Coulomb screening effects at their interfaces. These discoveries provide pathways for modulating correlated states, designing artificial polytype interfaces, and constructing functional heterostructures. In the future, by further regulating doping concentrations, pulsed laser irradiation, or localized electric fields, it may be possible to achieve more complex and patterned polytype interfacial structures within the same material system.

**Conclusion**

In summary, both vertical and lateral polytype heterostructures have been successfully realized within the low-hole-doped 1$T$-TaS$_2$ via thermally induced structural transitions. Raman spectroscopy, SKPM, and STM/STS measurements collectively identify the coexistence of the 1T- and 1H-phases and resolve their distinct electronic characteristics at the atomic scale. At the 1H/1T interface, a pronounced electronic transparency effect emerges, where the CDW modulation of the underlying 1T layer becomes visible through the metallic 1H-layer. By comparing 1T/1H and 1T/1T stacking configurations, we find that the metallic 1H-layer imposes Coulomb screening effect on the neighboring 1T-layer, suppressing intralayer electronic correlations and eliminating the formation of CDW domains. Overall, our results establish a robust approach for creating high-quality polytype heterostructures within a single crystal and provide a solid materials basis for future low-dimensional quantum devices enabled by phase-engineered architectures.

## Methods

**Sample preparation.** The high-quality 1$T$-(Ta$_{0.95}$Ti$_{0.5}$)S$_2$ single-crystal samples were grown by the chemical vapor transport (CVT) method with iodine as the transport agent. Ta (99.99%, Aladdin), Ti (99.99%, Aladdin), and S (99.99%, Aladdin) powders with a nominal mole ratio of 0.95:0.5:2 was weighed, mixed with 0.2 g of I$_2$, and placed into silicon quartz tubes. The doping has been uniformly and precisely controlled in high-quality samples. In addition, the sample quality was examined at multiple points using X-ray energy dispersive spectroscopy (EDS) with a scanning electron microscopy (SEM), which can give the average stoichiometry. The polytype heterostructures were synthesized by heating 1$T$-(Ta$_{0.95}$Ti$_{0.5}$)S$_2$ to 750 K in high vacuum (<10$^{-7}$ Torr) or in an argon purged glovebox. 1$T$-(Ta$_{0.95}$Ti$_{0.5}$)S$_2$ was held at 750 K for ~30 min, then brought down to room temperature.

**Scanning tunneling microscopy (STM).** High-quality Ti-doped 1$T$-TaS$_2$ crystals were cleaved at room temperature in ultrahigh vacuum at a base pressure of 2×10$^{-10}$ Torr, and directly transferred to the cryogen-free variable-temperature STM system (PanScan Freedom, RHK). Chemically etched W tips were used for STM measurement in constant-current mode. The STM tips were calibrated on a clean Ag(111) surface. Gwyddion was used for STM data analysis.

## Acknowledgments

This project was supported by the National Key R&D Program of China (MOST) (Grant No. 2023YFA1406500), the National Natural Science Foundation of China (NSFC) (No. 92477128, 92477205, 12374200, 11604063, 11974422, 12104504), the Strategic Priority Research Program (Chinese Academy of Sciences, CAS) (No. XDB30000000), and the Fundamental Research Funds for the Central Universities and the Research Funds of Renmin University of China (No. 21XNLG27). Y.Y. Geng was supported by the Outstanding Innovative Talents Cultivation Funded Programs 2023 of Renmin University of China. This paper is an outcome of "Two-dimensional anisotropic series of materials $FePd_{2+x}Te_2$: a structural modulation study from the atomic scale to the mesoscopic scale " (RUC25QSDL128), funded by the "Qiushi Academic-Dongliang" Talent Cultivation Project at Renmin University of China in 2025.

## Competing Interests

The authors declare no competing financial interests.

## Data Availability

The authors declare that the data supporting the findings of this study are available within the article and its Supplementary Information.

**References:**

[1] J. M. Carpinelli, H. H. Weitering, E. W. Plummer, and R. Stumpf, Direct observation of a surface charge density wave, Nature **381**, 398 (1996).

[2] W. Wang, B. Wang, Z. Gao, G. Tang, W. Lei, X. Zheng, H. Li, X. Ming, and C. Autieri, Charge density wave instability and pressure-induced superconductivity in bulk 1$T$-NbSe$_2$, Phys. Rev. B **102**, 155115 (2020).

[3] C. S. Lian, C. Si, and W. H. Duan, Unveiling charge-density wave, superconductivity, and their competitive nature in two-dimensional NbSe$_2$, Nano Lett. **18**, 2924 (2018).

[4] Y. Chen, W. Ruan, M. Wu, S. Tang, H. Ryu, H.-Z. Tsai, R. L. Lee, S. Kahn, F. Liou, C. Jia, O. R. Albertini, H. Xiong, T. Jia, Z. Liu, J. A. Sobota, A. Y. Liu, J. E. Moore, Z.-X. Shen, S. G. Louie, and S.-K. Mo, Strong correlations and orbital texture in single-layer 1$T$-TaSe$_2$, Nat. Phys. **16**, 218 (2020).

[5] J. M. Lu, O. Zheliuk, I. Leermakers, N. F. Q. Yuan, U. Zeitler, K. T. Law, and J. T. Ye, Evidence for two-dimensional Ising superconductivity in gated MoS$_2$, Science **350**, 1353 (2015).

[6] P. Wan, Y. Peng, J. Shen, X. Zhang, J. Wang, X. Luo, Z. Liu, T. Song, S. P. Parkin, and J. Wang, Orbital Fulde-Ferrell-Larkin-Ovchinnikov state in an Ising superconductor, Nature **619**, 46 (2023).

[7] K. F. Mak, C. Lee, J. Hone, J. Shan, and T. F. Heinz, Atomically thin MoS$_2$: A new direct-gap semiconductor, Phys. Rev. Lett. **105**, 136805 (2010).

[8] L. Ma, C. Ye, Y. Yu, X. F. Lu, X. Niu, S. Kim, D. Feng, D. Tomanek, Y.-W. Son, X. H. Chen, and Y. B. Zhang, A metallic mosaic phase and the origin of Mott-insulating state in 1$T$-TaS$_2$, Nat. Commun. **7**, 10956 (2016).

[9] B. Sipos, A. F. Kusmartseva, A. Akrap, H. Berger, L. Forró, and E. Tutiš, From Mott state to superconductivity in 1$T$-TaS$_2$, Nat. Mater. **7**, 960 (2008).

[10] M. Calandra, Phonon-assisted magnetic Mott-insulating state in the charge density wave phase of single-layer 1$T$-NbSe$_2$, Phys. Rev. Lett. **121**, 026401 (2018).

[11] K. S. Novoselov, A. Mishchenko, A. Carvalho, and A. H. Castro Neto, 2D materials and van der Waals heterostructures, Science **353**, aac9439 (2016).

[12] X. Huang, T. Wang, S. Miao, C. Wang, Z. Li, Z. Lian, T. Taniguchi, K. Watanabe, S. Okamoto, D. Xiao, S.-F. Shi, and Y.-T. Cui, Correlated insulating states at fractional fillings of the WS$_2$/WSe$_2$ moiré lattice, Nat. Phys. **17**, 715 (2021).

[13] S. K. Mahatha, J. Phillips, J. Corral-Sertal, D. Subires, A. Korshunov, A. Kar, J. Buck, F. Diekmann, G. Garbarino, Y. P. Ivanov, A. Chuvilin, D. Mondal, I. Vobornik, A. Bosak, K. Rossnagel, V. Pardo, A. O. Fumega, and S. Blanco-Canosa, Self-stacked 1T-1H layers in 6R-NbSeTe and the emergence of charge and magnetic correlations due to ligand disorder. ACS Nano **18**, 21052–21060 (2024).

[14] F. Lüpke, S. M. de Vasconcelos, B. A. Piot, M. H. Naik, J. R. Schaibley, J. Yan, D. G. Mandrus, K. Watanabe, T. Taniguchi, J. Zaanen, A. Rubio, X. Xu, and A. W. Tsen, Proximity-induced superconducting gap in the quantum spin Hall edge state of monolayer WTe$_2$, Nat. Phys. **16**, 526 (2020).

[15] A. K. Nayak, A. Steinbok, Y. Roet, J. Koo, G. Margalit, I. Feldman, A. Almoalem, A. Kanigel, G. A. Fiete, B. Yan, Y. Oreg, N. Avraham, and H. Beidenkopf, Evidence of topological boundary modes with

topological nodal-point superconductivity, Nat. Phys. **17**, 1413 (2021).

[16] K. Fan, H. Jin, B. Huang, G. Duan, R. Yu, Z.-Y. Liu, H.-N. Xia, L.-S. Liu, Y. Zhang, T. Xie, Q.-Y. Tang, G. Chen, W.-H. Zhang, F. C. Chen, X. Luo, W. J. Lu, Y. P. Sun, and Y.-S. Fu, Artificial superconducting Kondo lattice in a van der Waals heterostructure, Nat. Commun. **15**, 8797 (2024).

[17] A. Castellanos-Gomez, X. Duan, Z. Fei, H. R. Gutierrez, Y. Huang, X. Huang, J. Quereda, Q. Qian, E. Sutter, and P. Sutter, Van der Waals heterostructures, Nat. Rev. Methods Primers **2**, 58 (2022).

[18] X. Hong, J. Kim, S.-F. Shi, Y. Zhang, C. Jin, Y. Sun, S. Tongay, J. Wu, Y. Zhang, and F. Wang, Ultrafast charge transfer in atomically thin $MoS_2$/$WS_2$ heterostructures, Nat. Nanotechnol. **9**, 682 (2014).

[19] V. R. Policht, H. Mittenzwey, O. Dogadov, M. Katzer, A. Villa, Q. Li, B. Kaiser, A. M. Ross, F. Scotognella, X. Zhu, A. Knorr, M. Selig, G. Cerullo, and S. Dal Conte, Time-domain observation of interlayer exciton formation and thermalization in a $MoSe_2$/$WSe_2$ heterostructure, Nat. Commun. **14**, 7273 (2023).

[20] A. K. Geim and I. V. Grigorieva, Van der Waals heterostructures, Nature **499**, 419 (2013).

[21] L. L. Chang and L. Esaki, Semiconductor superlattices by MBE and their characterization, Prog. Cryst. Growth Charact. **2**, 3 (1979).

[22] Z. Zhou, F. Hou, X. Huang, G. Wang, Z. Fu, W. Liu, G. Yuan, X. Xi, J. Xu, J. Lin, and L. Gao, Stack growth of wafer-scale van der Waals superconductor heterostructures, Nature **621**, 499 (2023).

[23] A. Achari, J. Bekaert, V. Sreepal, A. Orekhov, P. Kumaravadivel, M. Kim, N. Gauquelin, P. B. Pillai, J. Verbeeck, F. M. Peeters, A. K. Geim, M. V. Milošević, and R. R. Nair, Alternating superconducting and charge density wave monolayers within bulk 6R-$TaS_2$, Nano Lett. **22**, 15 (2022).

[24] M. H. Fischer, F. Loder, and M. Sigrist, Superconductivity and local noncentrosymmetricity in crystal lattices, Phys. Rev. B **84**, 184533 (2011).

[25] H. Y. Dong, P. H. Sun, L. Lei, Y. Y. Geng, J. F. Guo, Y. Li, L. Huang, R. Xu, F. Pang, W. Ji, W. C. Zhou, Z. Liu, Z. Y. Lu, H. J. Gao, K. Liu, and Z. H. Cheng, Emergent electronic Kagome lattice in correlated charge-density-wave state of 1$T$-$TaS_2$, arXiv:2301.05885 (2023).

[26] H. Zhang, J. Guo, S. Mi, S. Li, and Z. Cheng, Kelvin probe force microscopy study on chiral superconductor 4Hb-$TaS_2$, Chin. J. Vac. Sci. Technol. **43**, 10 (2023).

[27] Y. Geng, H. Dong, R. Wang, J. Guo, S. Mi, L. Lei, Y. Li, L. Huang, F. Pang, R. Xu, W. Yu, H.-J. Gao, W. Ji, and Z. Cheng, Filling-dependent intertwined electronic and atomic orders in the flat-band state of 1$T$-$TaS_2$, ACS Nano **19**, 7784 (2025).

[28] Y. Y. Geng, L. Lei, H. Y. Dong, J. F. Guo, S. Mi, Y. Li, L. Huang, F. Pang, R. Xu, W. C. Zhou, Z. Liu, W. Ji, and Z. H. Cheng, Hysteretic electronic phase transitions in correlated charge-density-wave state of 1$T$-$TaS_2$, Phys. Rev. B **107**, 195401 (2023).

[29] Y. Geng, H. Dong, R. Wang, Z. Wang, J. Guo, S. Mi, Y. Li, F. Pang, R. Xu, L. Huang, H.-J. Gao, W. Ji, S. Wang, W. Zhou, and Z. Cheng, Real-space titration and manipulation of particle-like correlated electrons in doped Mott insulator, arXiv:2507.05756 (2025).

[30] J. J. Gao, J. G. Si, X. Luo, J. Yan, Z. Z. Jiang, W. Wang, Y. Y. Han, P. Tong, W. H. Song, X. B. Zhu, Q. J. Li, W. J. Lu, and Y. P. Sun, Origin of large magnetoresistance in candidate chiral superconductor 4Hb-$TaS_2$, Phys. Rev. B **102**, 075138 (2020).

[31] A. K. Nayak, A. Steinbok, Y. Roet, J. Koo, G. Margalit, I. Feldman, A. Almoalem, A. Kanigel, B. Yan, Y. Oreg, N. Avraham, and H. Beidenkopf, Evidence of topological boundary modes with topological nodal-point superconductivity, Nat. Phys. **17**, 1413 (2021).

[32] V. Vaňo, M. Amini, S. C. Ganguli, G. Chen, J. L. Lado, S. Kezilebieke, and P. Liljeroth, Artificial heavy fermions in a van der Waals heterostructure, Nature **599**, 582 (2021).

[33] A. Ribak, R. M. Skiff, M. Mograbi, P. K. Rout, M. H. Fischer, J. Ruhman, K. Chashka, Y. Dagan, and A. Kanigel, Chiral superconductivity in the alternate stacking compound 4Hb-TaS$_2$, Sci. Adv. **6**, eaax9480 (2020).

[34] I. Silber, S. Mathimalar, I. Mangel, A. K. Nayak, O. Green, N. Avraham, H. Beidenkopf, I. Feldman, A. Klein, M. Goldstein, A. Banerjee, E. Sela, and Y. Dagan, Two-component nematic superconductivity in 4Hb-TaS$_2$, Nat. Commun. **15**, 824 (2024).

[35] J. Bang, B. Lee, H. Yang, S. Kim, D. Wulferding, and D. Cho, Charge-ordered phases in the hole-doped triangular Mott insulator 4Hb-TaS$_2$, Phys. Rev. B **109**, 195170 (2024)

[36] A. K. Nayak, A. Steinbok, Y. Roet, J. Koo, I. Feldman, A. Almoalem, A. Kanigel, B. Yan, A. Rosch, N. Avraham, and H. Beidenkopf, First-order quantum phase transition in hybrid metal-Mott insulator 4Hb-TaS$_2$, Proc. Natl. Acad. Sci. U.S.A. **120**, e2304274120 (2023).

[37] L. Yang, S. Hu, G. Hu, W. Zhou, Y. Zhang, T. Dai, Y. Zhang, J. Zhang, C. Liu, J.-O. Wang, J. Qiao, Z. Li, Y. Shao, and X. Wu, Tuning the crystalline phase transition temperature of 1$T$-TaS$_2$ via surface oxidation, Adv. Mater. Interfaces **12**, 12 (2025).

[38] Y. Chen, Y.-X. Dai, Y. Zhang, C. Zhang, L. Zhou, L. Jia, W. Wang, X. Han, H. Yang, L. Liu, C. Si, Q.-F. Sun, and Y. Wang, Nanoscale polymorph engineering of metal-correlated insulator junctions in monolayer NbSe$_2$, ACS Nano **19**, 14808 (2025).

[39] Z. Wang, Y.-Y. Sun, I. Abdelwahab, L. Cao, W. Yu, H. Ju, J. Zhu, W. Fu, L. Chu, H. Xu, and K. P. Loh, Surface-limited superconducting phase transition on 1$T$-TaS$_2$, ACS Nano **12**, 12619 (2018).

[40] Y. Geng, J. Guo, F. Meng, M. Wang, S. Mi, L. Huang, R. Xu, F. Pang, K. Liu, S. Wang, H.-J. Gao, W. Zhou, W. Ji, H. Lei, and Z. Cheng, Correlated electrons of the flat band in charge-density-wave state of 4Hb-TaSe$_x$S$_{2-x}$, Phys. Rev. B **110**, 115107 (2024).